# Dark matter and cosmic structure

*Carlos S. Frenk*[1, *] *and Simon D. M. White*[2]

We review the current standard model for the evolution of cosmic structure, tracing its development over the last forty years and focussing specifically on the role played by numerical simulations and on aspects related to the nature of dark matter.

## 1 Preamble

The development of a standard model of cosmological structure formation over the last forty years must count as one of the great success stories in Physics. The model describes the geometry and material content of the Universe, explaining how structure - galaxies of various sizes and types, groups and clusters of galaxies, the entire cosmic web of filaments and voids - emerged from a hot and near-uniform Big Bang. This astonishing richness appears to have grown from quantum zero-point fluctuations produced during a period of cosmic inflation occurring very soon after the beginning of our Universe. Over the subsequent 13.7 billion years, these small density perturbations were amplified by the relentless action of gravitational forces due predominantly to dark matter, but illuminated by complex astrophysical phenomena involving ordinary matter. In the standard model, the dark matter, which makes up five sixths of all matter, is hypothesised to be a weakly interacting non-baryonic elementary particle, created in the early stages of cosmic evolution.

In this article we will review the standard model of cosmic structure formation, focusing on the nonlinear processes that transformed the near-uniform universe that emerged from the Big Bang into the highly structured world we observe today. We will highlight the critical role that computer simulations have played in understanding this transformation, emphasizing gravitational phenomena associated with the dark matter, and only rather briefly discussing the crucially important, but more complex, processes that shape the directly observable baryonic components. Rather than presenting a complete and thorough review of the literature, our intention is to provide an account of the main ideas and advances that have shaped the subject. We begin by presenting in Table 1 a chronological listing of the landmark developments that have driven this remarkable story.

## 2 Prehistory

In 1933 Zwicky published unambiguous evidence for dark matter in the Coma galaxy cluster [1]; in 1939 Babcock's rotation curve for the Andromeda Nebula [2] indicated that much of its mass is at large radius; in 1959 Kahn & Woltjer argued that the total mass of the Milky Way and Andromeda galaxies must be much larger than their stellar mass in order to explain why they are currently approaching each other [3]. Nevertheless, the idea that galaxies and galaxy clusters are embedded in massive dark matter halos was not widely discussed until the 1970's. This changed with the influential, observationally based papers of Ostriker, Peebles and Yahil [4] and Einasto, Kaasik, Saar and Chernin [5] and with Ostriker and Peebles' theoretical argument that massive halos are required to stabilize spiral galaxy disks [6]. Within four years, a hierarchically merging population of dark matter halos formed the basis for the galaxy formation scenario proposed by White and Rees [7], providing the gravitational potential wells within which gas cools and condenses to form galaxies. Although massive neutrinos had already been suggested as a dark matter candidate by Cowsik and McClelland [8,9] and Szalay and Marx [10], White and Rees considered an early population of low-mass stars as the most plausible identification for the dark matter in their theory.

The evolution of the halo population in this theory was based on Press and Schechter's [11] analytic model for the growth of cosmic structure from a Gaussian initial density field. To test this model, Press and Schechter carried out computer N-body simulations. This was the first time numerical experiments were used for quantitative exploration of nonlinear structure formation in an expanding universe.

* Corresponding author    E-mail: c.s.frenk@durham.ac.uk
[1] Institute for Computational Cosmology, Unversity of Durham, U.K.
[2] Max-Planck Institute for Astrophysics, Garching, Germany

**Table 1** A timeline of key developments

---

*1970-1980*      **Prehistory**

Dark matter (DM) halos are proposed to surround all galaxies

Massive neutrinos are suggested as a dark matter candidate

Large-scale structure is characterized through the measurement of galaxy correlation functions

First analytic models and computer simulations are built for nonlinear structure formation in an expanding universe

Simulations from idealized initial conditions contrast the "isothermal" case where structure grows by hierarchical merging with the "adiabatic" case where galaxies form by fragmentation of large-scale structure

*1980-83*      **The breakthrough years**

The CfA redshift survey provides a first representative picture of the cosmic web

A claimed measurement of a 30eV mass for $\nu_e$ galvanizes research into elementary particle dark matter

Quantum fluctuations during an early inflationary era are suggested as a physical model for the initial generation of density fluctuations

Improved calculations of early linear evolution provide accurate initial conditions for late-time growth, dividing particle candidates into Hot, Warm and Cold Dark Matter (HDM, WDM, CDM respectively)

Simulations from these initial conditions exclude HDM (and hence all known particle species) for the bulk of the DM

*1983-2002*      **Establishment of the standard model**

Simulations from CDM initial conditions produce large-scale structure resembling that seen in redshift surveys

Galaxies are shown to trace the DM distribution in a biased way depending strongly on galaxy type and on epoch

The first direct detection experiments and the first searches for DM annihilation radiation are set up

The COBE measurement of temperature fluctuations in the cosmic microwave background (CMB) indicates an amplitude too small for a universe without particle DM, but too large for a universe with a closure density of such DM

Large-scale correlations of galaxies and the baryon fraction of rich clusters prove inconsistent with a high-density universe

Simulations establish the detailed structure expected for DM halos in a CDM universe

Supernova measurements rule out a high density universe and strongly suggest accelerated expansion

Boomerang and Maxima convincingly localize the first acoustic peak in the CMB, showing our universe to be flat

WMAP convincingly measures the second acoustic peak also, excluding many alternatives to the standard model and giving precise estimates of the baryon and dark matter densities

*2002-present*      **Pushing to new frontiers**

Simulations explore the predictions of the standard model at higher precision and on both larger and smaller scales

Gravitational lensing measurements check the detailed structure predicted for dark halos

Large redshift surveys give precise measures of galaxy biasing, and of the Gaussianity and power spectrum of the initial conditions, detecting weak baryon-induced features

Simulations and observations explore constraints on the nature of the dark matter based on the inner structure of dark halos and on structure in the intergalactic medium

Both direct detection experiments and the Fermi satellite begin to exclude parts of the parameter space expected for WIMPs

---

 

Soon after, White's simulation of the assembly of the Coma cluster [12] provided the first detailed study of the hierarchical growth of an individual object from expanding initial conditions. Systematic numerical studies of the growth of large-scale structure were carried out by Aarseth, Gott and Turner [13] (also [14, 15]), by Efstathiou [16] and by Efstathiou and Eastwood [17]. This work aimed at understanding whether gravitational evolution could explain the quantitative properties of the galaxy distribution as revealed by the clustering studies of Peebles and collaborators [18] and the group catalogue of Gott and Turner [19].

The lack of an agreed identification of the gravitationally dominant component of the universe, and of a physical theory for the origin of cosmic structure, meant that all these calculations started from highly idealized initial conditions. Discussion centred on the contrast between "adiabatic" models, in which early linear evolution erased all structure smaller than the Silk damping length [20] or the neutrino free-streaming length [21], and "isothermal" models where small-scale fluctuations survive to late times. In the former case the first nonlinear objects are superclusters which must fragment to form galaxies [22]; in the latter case the first objects are much smaller and galaxies build up through hierarchical clustering [18].

Early structure formation N-body simulations represented high-redshift density fluctuations in the isothermal scenario by discreteness noise in the initial distribution of simulation particles. The direct N-body integrators of the 1970's could follow ~ 1000 particles within an expanding spherical volume ($N = 4000$ in two simulations by Aarseth, Gott and Turner [13]) so small particle numbers and edge effects seriously limited the conclusions which could be drawn. Simulations of adiabatic models in 3D were carried out somewhat later, using initial conditions where a regular particle grid was perturbed by a superposition of linear growing modes with a sharp short-wavelength cut-off [23–25]. Use of Fourier force evaluation techniques substantially increased the number of particles that could be followed ($N = 32,768$ in [23], $N = 10^6$ in [25]) and also eliminated edge effects through the use of periodic boundary conditions. (Fourier calculation of long-range forces had already allowed Efstathiou and Eastwood [17] to simulate the isothermal scenario with $N = 20,000$ in a periodic "box".) Nonlinear growth was found to produce near power-law autocorrelation functions at certain times in both scenarios, but large-scale structure was clearly more coherent in the adiabatic case. Quantitative comparison with observation was hampered by the schematic initial conditions, the relatively small numbers of particles used, and the uncertain relation between real galaxies and simulation particles.

These and almost all later simulations have used Newtonian rather than relativistic gravity. This is an excellent approximation since linear structure growth is identical in the matter dominated regime in the two theories, and nonlinear large-scale structure induces velocities far below the speed of light. In future, however, relativistic corrections may eventually be needed [26]

## 3 The breakthrough years

Today's standard model of cosmology arose through the confluence of three distinct disciplines: particle physics, astronomy and computing. In the early 1980s, new developments in each of these areas coalesced to establish the physical foundations and methodology of what, twenty years later, would become the standard model.

From particle physics came two fundamental new ideas. The first was Guth and Linde's theory of cosmic inflation [27, 28] and the realisation that it could give rise to quantum fluctuations that might seed the universe with adiabatic, scale invariant density perturbations [29–32]. The second was the proposal that the dark matter (DM) could be composed of non-baryonic particles. This idea took centre stage after Lubimov et al. [33] measured a ~ 30eV mass for the electron neutrino, enough to provide the critical density needed to close the universe [8]. Such light particles would remain relativistic until relatively late times, but it was soon recognized that other (hypothetical) particles which were more massive and decoupled earlier could also be dark matter. This led to the convenient classification of candidate dark matter particles into three families: hot, warm and cold dark matter, names that reflect their typical velocities at some early time, for example at the epoch of recombination [21]. Light neutrinos are the prototype for hot dark matter (HDM) and are the only DM candidates that are proven to exist from earth-bound experiments; a supersymmetric particle [34] or an axion [35] emerged as plausible candidates for cold dark matter (CDM); a nonstandard gravitino was the initially suggested candidate for warm dark matter (WDM) [36], but more recently, a sterile neutrino has seemed a more attractive possibility [37].

For the first time, there was a plausible physical mechanism to generate small density irregularities in the very early universe and several concrete proposals for the identity of the dark matter which would drive their late-time amplification. Thus, the ingredients were in place for detailed linear calculations of the evolution of structure in the universe, from its initial generation until the formation of the first nonlinear objects. Even though the experimental result of Lubimov et al., so influential at the time, turned out to be wrong, these revolutionary and initially controversial ideas form the basis of modern cosmological thinking.

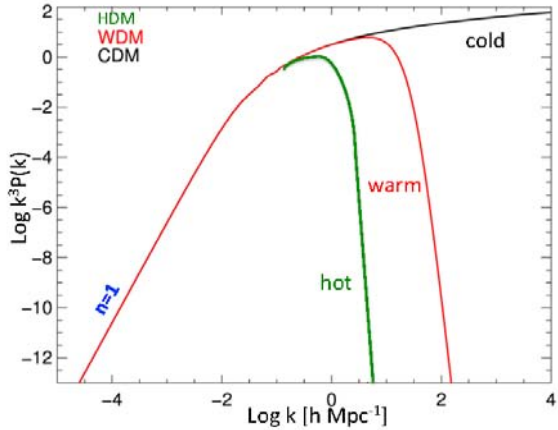

Inflation seeds the universe with nearly scale invariant, adiabatic density perturbations of very small (and tunable) amplitude and a simple power-law power spectrum, $P(k) \propto k^n$, where $n$ is close to, but smaller than unity. As the universe expands, the growth of these perturbations is regulated initially by the dominant radiation component and later by the dark matter. During the radiation era the growth of matter perturbations slows as they are encompassed by the cosmological particle horizon (the "Meszaros effect" [38]). This imprints a characteristic scale corresponding to the horizon at the time the universe became matter dominated; on smaller scales (larger $k$) the power spectrum bends to $P(k) \propto k^{n-3}$. In addition, dark matter fluctuations are washed out by random thermal motions below a free-streaming scale corresponding to the typical comoving distance that a particle travels in the age of the universe[1]. This varies approximately inversely with particle mass, $m_X$, i.e. $\lambda_{fs} \propto m_X^{-1}$.

The resulting linear theory power spectra at late times (say, at recombination) are illustrated for hot, warm and cold dark matter in Fig. 1. For HDM with $m_X \sim 30$ eV, the characteristic mass corresponding to the free-streaming length is roughly that of a large galaxy cluster; for WDM with $m_X \sim 2$ keV, it corresponds to the dark halo of a dwarf galaxy; while for CDM with $m_X \sim 100$ GeV, it corresponds to the Earth's mass. Thus, the smallest structure which can form directly is radically different in each case: for HDM, superclusters form first and must fragment to make galaxies; for WDM and CDM, small objects form first and grow by merging and accretion to make bigger systems. Dark matter objects significantly smaller than galaxies can form in CDM, but not in WDM.

The proposal of inflationary fluctuation generation and the rise of interest in particle dark matter coincided with the publication of the first extensive 3D survey of galaxies, the CfA redshift survey [39]. Lilliputian by today's standards, this survey gave the first clear picture of the richness of the large-scale distribution of galaxies, offering a glimpse of what would later be called the "cosmic web" [40]. The CfA survey called for an explanation of the origin of this web. The ideas coming from particle physics offered the enticing prospect of understanding the cosmic large-scale structure as a consequence of fundamental physics. All that was needed were the tools to calculate how the linear initial conditions would be reflected in today's observable, highly nonlinear universe.

As discussed in § 2, N-body simulations had been used in previous years to study the growth of nonlinear structure in an expanding universe, and the size of feasible calculations was increasing rapidly through developments in both hardware and algorithms. The two critical developments needed before simulations could link the early universe to the CfA survey were algorithms to represent the growing linear power spectra of Fig. 1 in the N-body initial conditions, and an understanding of where to place the observed "CfA" galaxies in the final mass distribution. The first problem is purely technical and could be solved by careful application of noise-suppression and Fourier methods within "boxes" representing the fundamental cube of a triply periodic universe [41]. The second problem involves significant astrophysical uncertainties about how galaxies form and will be discussed in detail in the next section. In the HDM case, however, it was possible to get a major result without a complete solution to this issue.

---

[1]  Axions are never in thermal equilibrium and, as soon as they acquire their (very low) mass, they behave like cold dark matter.

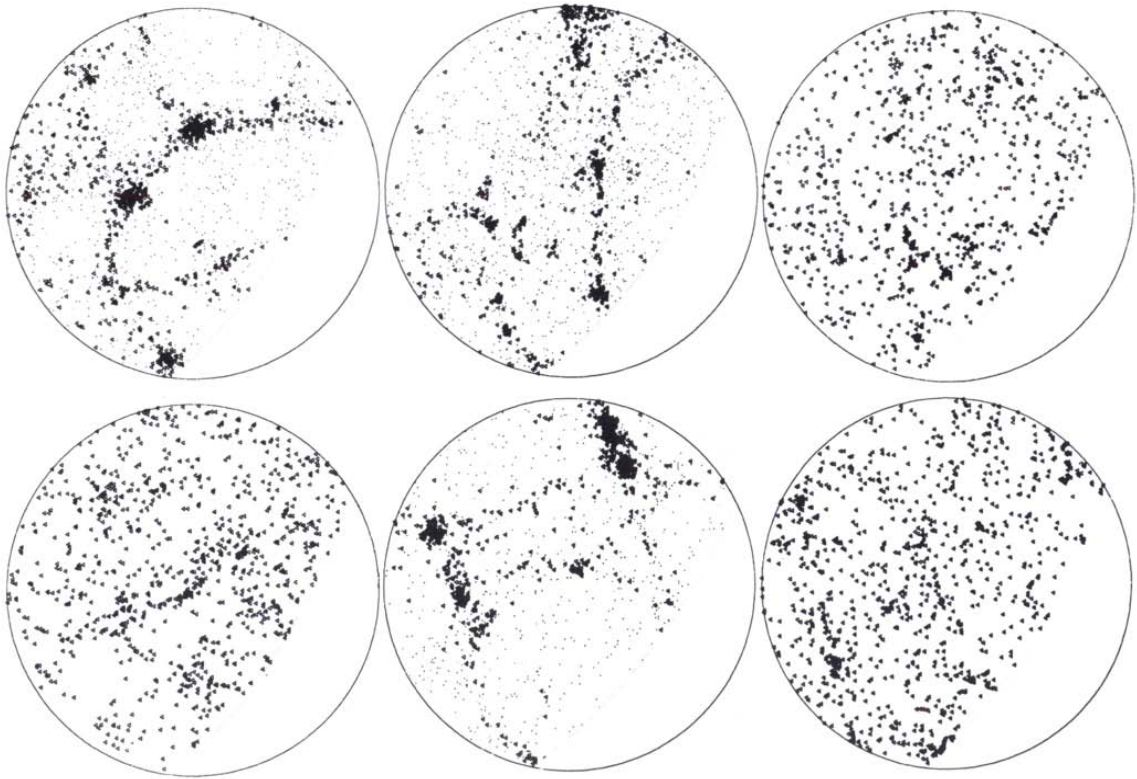

**Figure 2** Plots of the sky distribution of "galaxies" in N-body simulations of three hot dark matter universes and two cold dark matter universes, compared to the galaxy distribution in the CfA redshift survey. The HDM models (middle top, middle-bottom and right-bottom panels) have $\Omega_m = 1$ and the possible locations of galaxies are marked as triangles. The CDM models (left top and left bottom panels) assume $\Omega_m = 0.2$. In the HDM case galaxies were assumed to form only in the large-scale structure where the dark matter has locally collapsed; in the CDM case galaxies are simply assumed to trace the mass.

The idea that the dark matter could consist of light neutrinos with the mass measured by Lubimov et al. [33] was immensely attractive: neutrinos were known to exist and a mass of ∼ 30 eV would imply the simplest possible cosmological model, a universe with the critical density, $\Omega = 1$ [2]. Unfortunately, simulations of structure growth from the initial conditions predicted by inflationary models for such a neutrino-dominated universe demonstrated that these ideas lead to a universe quite unlike the one we see around us

---

[2] In the 1980s, a matter-dominated universe with the critical density was regarded as a theoretical imperative even though many observational astronomers favoured an open universe.

[42]. As Fig. 2 shows, the free-streaming cut-off in the power spectrum imprints a large and well-defined scale on the galaxy distribution which is clearly not present in the real universe, as revealed by the CfA redshift survey. In such a HDM universe, galaxies form only in supercluster regions where the matter distribution has locally collapsed; these regions have such a large coherence scale that the pattern imprinted on the galaxy population at formation is already much more clustered than that seen in the CfA survey. This severe discrepancy resulted in an immediate loss of interest in HDM models, even though direct mass measurements did not exclude the last HDM candidate known to exist, the $\tau$ neutrino, for another 20 years [43].

Although disappointing from the point of view of explaining the large-scale structure of the CfA survey, the exclusion of a neutrino-dominated universe was a major success for the then new cosmological methodology: a light neutrino mass in the $\sim 30$ eV range was conclusively ruled out by a combination of N-body simulations and astronomical observations. Attention then shifted to the alternative possibility of a universe dominated by cold dark matter. The first simulations of structure formation in such a universe [44] (hereafter DEFW), illustrated in Fig. 2, revealed that this hypothesis gave far better results when compared to the CfA data. The plots shown correspond to a model with matter density $\Omega_m = 0.2$ either in an open universe or in a flat universe with $\Omega_\Lambda = 0.8$ (simulations in the two cosmologies were indistinguishable at the resolution of DEFW). The flat case was very close to today's standard $\Lambda$CDM model. In these low-density simulations, galaxies were assigned to random dark matter particles and the fluctuation amplitude was set so that the slope of their two-point correlation function roughly matched that observed. The resemblance to the CfA galaxy distribution, and the natural framework for galaxy formation that they provide [45], encouraged further exploration of this class of models, although the absence of any physical motivation or empirical evidence for a cosmological constant channeled attention to CDM variants with $\Omega_m = 1$.

Although today's standard model of cosmology is often taken for granted, it is rooted in ideas that were once revolutionary and were not easily accepted by the scientific community. On the contrary, the early days were marked by profound scepticism and outright opposition by many. It is worth remembering this when criticizing current radical ideas, for example large-scale modifications of the laws of gravity, or the addition of new dimensions to space-time.

# 4 Putting in the galaxies

Simulations of the evolution of cosmic large-scale structure use N-body methods to follow the gravitationally dominant dark matter component. Our most precise observational information about large-scale structure comes, however, from galaxy redshift surveys, beginning with the 2400 galaxy CfA survey [39] and continuing to the over $1.5 \times 10^6$ galaxies with redshifts in the current Sloan Digital Sky Survey (SDSS) catalogues[3]. Since stars in galaxies account for only a few percent of the matter density of the universe and galaxies form at special points of the overall mass distribution, namely at the centres of dark matter halos, a longstanding issue has been the appropriate way to compare simulated mass distributions with observed galaxy distributions.

Early simulations simply assigned "galaxies" to randomly chosen simulation particles and compared their clustering statistics directly with those of observed galaxies. This assumption had no physical basis and even at the time it was known to be incorrect, since different types of galaxy had already been shown to cluster differently [46]. Particularly marked differences are predicted between the distributions of dark matter and of galaxies in neutrino-dominated universes since the large regions between superclusters contain large amounts of dark matter but no collapsed objects, and hence no galaxies (see Fig. 2). Effects are more subtle in Warm or Cold Dark Matter universes but are still substantial. Rare, massive objects like galaxy clusters form from unusually dense regions of the early universe, and these are strongly clustered even *before* nonlinear structure forms [47]. Clustering today reflects both this initial pattern and the structure induced by gravitational growth, and is stronger than that of the dark matter, which reflects gravitational growth alone. On large scales, cluster number fluctuations are a factor $b$ (the "bias") times larger than mass fluctuations, where $b$ is independent of spatial scale but strongly increasing with cluster mass. Bardeen and collaborators [48] showed that the initial clustering pattern is well represented by assuming clusters to correspond to high peaks of a suitably smoothed version of the linear density fluctuation field, and derived many useful formulae for the standard case where this field is assumed to be Gaussian.

In the first set of simulations of structure growth from CDM initial conditions, DEFW compared the CfA galaxy distribution with predictions for a high-density (Einstein-de Sitter) universe, where galaxies were taken to corre-spond to high peaks of the smoothed initial density field, and with predictions for low-density universes (see Fig. 2) where galaxies were taken to be distributed in the same way as the dark matter. They found that the observed spatial clustering and dynamically induced motions of galaxies could be matched in both cases; a higher mean cosmic matter density can be compensated by biasing the observed amplitude of galaxy clustering high relative to the mass. This degeneracy requires $b$, the large-scale bias, and $\Omega_m$ the current matter density of the universe to satisfy $\Omega_m^{0.6}/b \sim 0.4$.

---

[3] http://www.sdss3.org

In the 1980's a flat universe was considered the natural outcome of inflation and a cosmological constant was considered unnatural because of the low implied energy-scale, so a biased Einstein-de Sitter model (called "standard" Cold Dark Matter or SCDM at the time) became the preferred choice and was extensively investigated using N-body simulations [49–54]. Nevertheless, by the early 1990's it had become clear that galaxy correlations are stronger on large scales than predicted by SCDM [55] and that the baryon fraction in rich clusters is inconsistent with standard cosmic nucleosynthesis in a high density universe [56]. Thus, alternatives to SCDM were tried throughout the decade [57]. The transition to ΛCDM was finally forced by the exclusion of an Einstein-de Sitter expansion history by supernova data [58, 59] and the demonstration that the universe is flat through localization of the first peak in the spectrum of microwave background fluctuations [60, 61]. Although the strong biasing of SCDM does not occur in today's universe, it is, however, present at high redshift ($z \gtrsim 1$) and is now an integral part of interpreting observations of the high-redshift universe.

A conceptually useful way of modelling the bias of galaxies relative to dark matter comes from an extension of the Press-Schechter model for the abundance of halos [62–65]. In its simplest and most useful form, this EPS theory implies that the assembly histories of halos of given mass (and thus the properties of the galaxies within them) are independent of each halo's larger scale environment [66]. In this approximation, the galaxy distribution can be fully characterised by specifying separately the distribution of halos as a function of mass, position and velocity, and the distribution of galaxies within halos of given mass. The former can be inferred with high accuracy from N-body simulations of representative cosmological volumes. However, the large-scale bias of halos can also be inferred more simply from a variant of the Press-Schechter argument developed by Cole and Kaiser [67]. For halo mass $M$ and redshift $z$, this gives

$$b(M, z) = 1 + [(\delta_c/\sigma(M, z))^2 - 1]/\delta_c, \qquad (1)$$

where $\sigma(M, z)$ is the *rms* linear fluctuation at redshift $z$ in fixed radius spheres containing, on average, mass $M$,

and $\delta_c = 1.68$ is the extrapolated linear overdensity at which a spherical perturbation collapses to zero radius in an Einstein-de Sitter universe. This formula was rederived from EPS theory and was extended to nonlinear scales by Mo and White [68] who also tested it directly against N-body simulations. It works well for massive halos ($\delta_c/\sigma > 1$) which are positively biased ($b > 1$) but underestimates $b$ for low-mass halos [69, 70]. A possible explanation for this may come from the work of Sheth, Mo and Tormen [71] who extended EPS theory to assume ellipsoidal rather than spherical collapse, resulting in improved fits to both the abundance and the clustering of simulated halos.

The distribution of galaxies within halos of given mass can be obtained either from a purely statistical model, whose structure and parameters must then be set to fit observational data (a Halo Occupation Distribution or HOD; [72–76]) or through direct modelling of galaxy formation during the assembly of halos [72, 77]). The precision achievable by either method is ultimately limited by the fact that the bias of dark matter halos depends not only on their mass, but also on their formation time, concentration, spin, substructure content and shape, all of which are also expected to influence their galaxy content. This effect is known as "assembly bias". Although it is empirically well established from large cosmological simulations, it is not yet well understood theoretically [78–84].

The limitation imposed by assembly bias can only be fully circumvented by following the formation and evolution of the galaxies within each dark halo in the simulated volume. It might seem that this would best be accomplished by direct inclusion of the baryonic component and all the astrophysical processes affecting it. Unfortunately, this is far beyond current computational capabilities. The difficulty is that the relevant baryonic processes are not only much more diverse and complex than the purely gravitational processes which affect the dark matter, but are also sensitive to much smaller length- and time-scales than can be followed in a cosmological simulation (related, for example, to star formation and evolution or black hole accretion). Although this is currently a very active area of computational astrophysics and progress is rapid, such hydrodynamic galaxy formation simulations are still far from reproducing basic properties of the galaxy population such as their abundance as a function of stellar mass or their spatial clustering [85–88].

An alternative approach takes the assembly history and structure of each dark halo from a large cosmological simulation and calculates the growth of galaxies within it using simplified phenomenological treatments of baryonic processes. The latter are typically based on physical insights gained from simulations of individual systems and from observation [89–91]. Uncertain parameters such as the ef-ficiencies of star formation, of black hole growth or of feedback from stars and black holes can be adjusted to reproduce the properties of the observed galaxy distribution, thus providing a practical and empirical route to measure these efficiencies and their dependence on galaxy mass and epoch. In practice, the flexibility and computational efficiency of this semi-analytic simulation technique allow extensive experimentation with the form and parameters of the various models for baryonic processes.

The earliest simulations of this type followed galaxy formation within halo merger trees [92–94] but with increasing simulation resolution it became possible to identify dark matter subhalos within individual halos and to assign their positions and velocities to the corresponding "satellite" galaxies [95–97]. Even at the highest resolution, however, convergent results for the galaxy distribution in clusters, in particular, for their number density as a function of radius, require keeping track of "orphan" galaxies. These are galaxies that are required to survive, at least temporarily, despite the tidal disruption of their simulated dark matter subhalos [95,98].

Recent semi-analytic galaxy formation simulations follow the evolution of millions of galaxies throughout volumes comparable to those of new very large redshift surveys like the SDSS. In particular, several generations of publicly available models based on the Millennium Simulatio [96, 98–101] have produced ever closer matches to the observed galaxy population and have been widely used by the community[4]. This success reflects the fact that simulations of this kind make it possible to construct mock surveys where the simulated galaxy population is "observed" with a virtual telescope to produce a sample in which galaxy properties and the large-scale structure can be compared directly with those observed in real surveys (see Fig.3). Such comparisons can be used to test the effectiveness of observational procedures for identifying galaxy groups and clusters and for measuring their masses [102–104]. They also brings to light both inadequacies in the galaxy formation modelling [96, 98] and, potentially, discrepancies with the underlying ΛCDM paradigm. For example, one can test for deviations from the Gaussian statistics assumed in the initial conditions (e.g. [105]) , or for deviations between the mean mass profiles of galaxy clusters, as measured directly by gravitational lensing, and the predictions of the theory [106,107].

# 5 Dark matter halos

Dark matter halos are the fundamental nonlinear units of cosmic structure and galaxies condense in their cores. As a result, halos have a special status in cosmology. Understanding their basic properties – formation histories, internal structure and abundance – is an important step in devising astronomical tests of the ΛCDM paradigm and in exploring how galaxies form. Halos are also the prime hunting ground for dark matter detection. Direct detection experiments target dark matter particles at the Earth's position within our own Galactic halo; indirect detection experiments target the radiation from decaying or annihilating dark matter particles, and this is also strongly affected by the structure of halos.

In this section we review five important aspects of dark matter halos: formation histories, abundance as a function of mass (the mass function), density and velocity structure, shapes and the properties of substructures. For high-resolution results, we will rely on recent simulations carried out by two groups: the Aquarius and Phoenix projects by the Virgo Consortium [112,113] and the Via Lactea and GHALO simulations [114,115]. Aquarius, Via Lactea and GHALO are N-body simulations of halos similar in mass to that of the Milky Way, Phoenix of halos about $10^3$ times more massive. For statistical results on the halo population we will rely primarily on the Millennium simulation series (which consists of the Millennium, Millennium-II and Millennium-XXL simulations [96,116,117]).

Images of an Aquarius and a Phoenix halo are displayed in Fig. 4 and illustrate some of the main characteristics that we will discuss in the remainder of this section: (i) the mass distribution is centrally concentrated and strongly aspherical; (ii) a large number of substructures are present, particularly in the outer parts.

## 5.1 Halo formation

Whether the dark matter is cold or warm, halos form in a hierarchical manner provided fluctuations were originally generated with a spectrum similar to that predicted by inflationary models, $P(k) \propto k^n$ with $n \approx 1$. Small structures are the first to become nonlinear (that is, to decouple from the universal expansion, collapse and reach a dynamical state near virial equilibrium); larger structures form subsequently by mergers of pre-existing halos and by accretion of diffuse dark matter that has never been part of a nonlinear object. Halos typically form "inside out", with a strongly bound core collapsing initially and material being gradually added on less bound orbits; only "major mergers" of similar mass halos lead to near-complete mixing of old and new material. Such hierarchical growth occurs for any approximately Gaussian initial conditions in which the fluc-tuation amplitude increases monotonically with decreasing scale.

---

[4]  http://www.mpa-garching.mpg.de/millennium. This website also lists the more than 500 papers that have been published to date using the Millennium Simulation data.

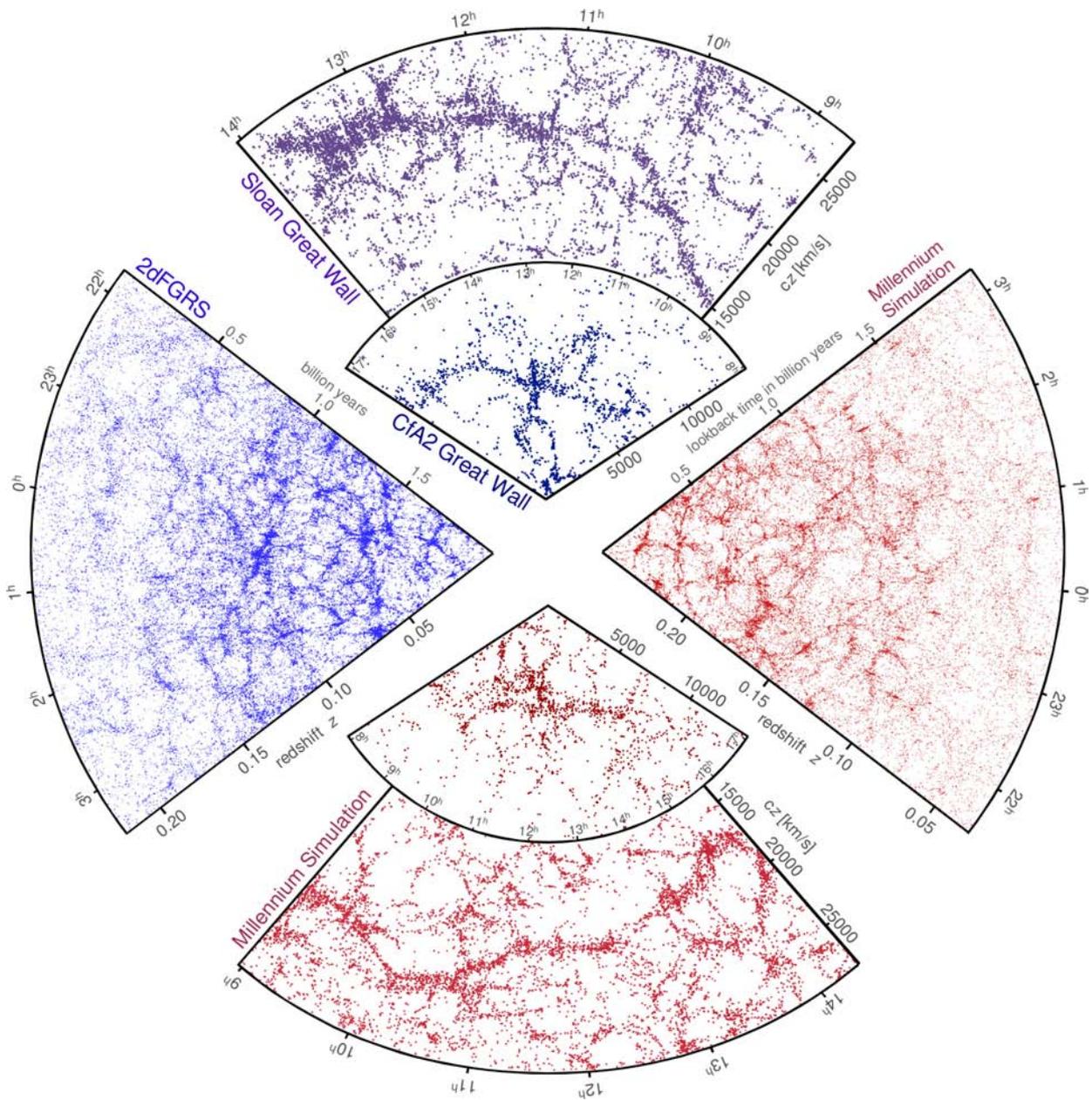

**Figure 3** The galaxy distribution in redshift surveys and in mock catalogues constructed from the Millennium Simulation. The small slice at the top shows the CfA2 'Great Wall' [108], with the Coma cluster at the centre. Drawn to the same scale is a small section of the SDSS, with the even larger 'Sloan Great Wall' [109]. The cone on the left shows one half of the 2-degree galaxy redshift survey (the 2dFGRS; [110]) . The cones at the bottom and on the right correspond to mock galaxy surveys with similar geometries and magnitude limits constructed by applying semi-analytic galaxy formation simulation methods to the halo/subhalo assembly trees of the Millennium Simulation [Adapted from [111]].

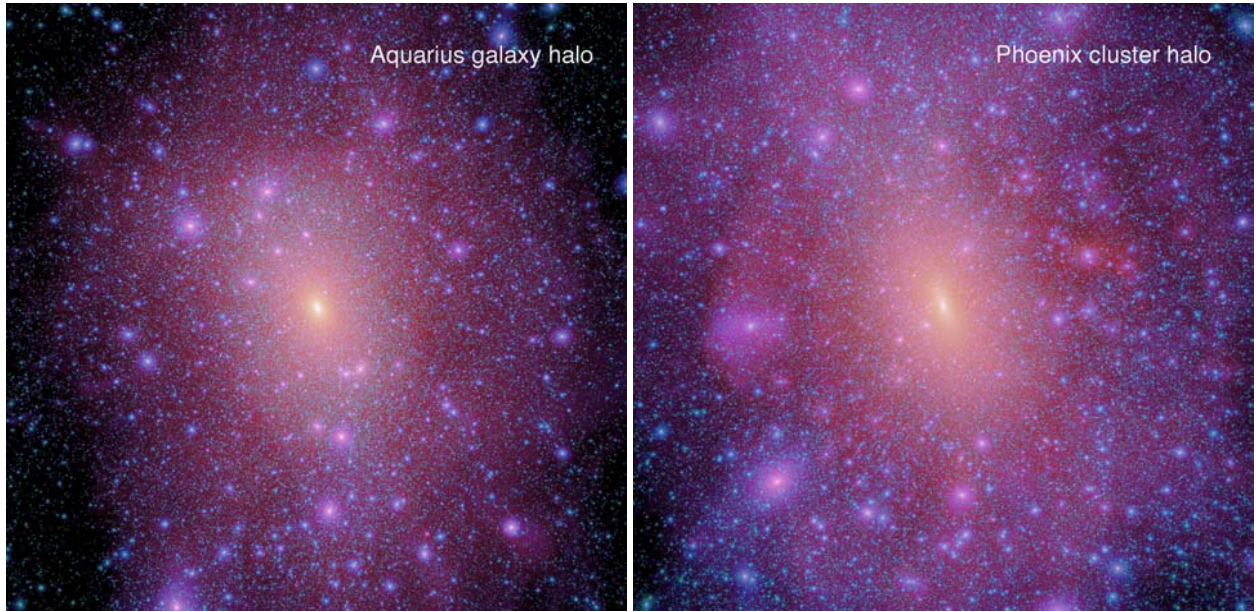

**Figure 4** Images of a galaxy (left) and a rich cluster (right) dark matter halo from the Aquarius and Phoenix simulation projects. The two images show the projected dark matter density at $z = 0$, in a box of side $1.07\,\mathrm{Mpc}$ for the galactic halo and $7.14\,\mathrm{Mpc}$ for the cluster halo (roughly 4 times $r_{200}$). The brightness of each image pixel is proportional to the logarithm of the square of the dark matter density projected along the line-of-sight, and the hue encodes the density-weighted average of velocity dispersion along the line-of-sight. The two images have the same number of resolution elements within the radius $R_{200}$. [Adapted from [118] and [113].]

A full analytic description of the development of such dissipationless hierarchical clustering came in the early 1990's with extensions of the original Press-Schechter model [11] based on excursion set theory [62–65]. In particular, this description provided formulae for halo merger rates and formation time distributions, as well as algorithms for producing Monte Carlo samples of halo assembly histories. All this machinery was tested against simulations of hierarchical clustering and found to fit well [65,119].

Modern large-volume and high-resolution simulations have quantified the statistics of halo mergers in the $\Lambda$CDM cosmology over a wide range of halo masses, providing accurate fitting formulae for the mean merger rate per halo over the mass range $10^{10} - 10^{15}\,\mathrm{M_\odot}$ and the redshift range $0 < z < 15$ [120]. If we define a formation epoch as the time when half the final halo mass was in a single object, then in the $\Lambda$CDM model halos of mass $10^8, 10^{10}, 10^{12}$ and $10^{14}\,\mathrm{M_\odot}$ have median formation redshifts of 2.35, 1.75, 1.17 and 0.65 respectively.

At the resolution of the Millennium-II Simulation, halos are resolved down to a mass of $2 \times 10^8\,\mathrm{M_\odot}$. At $z = 0$ such halos contain 60% of the total mass in the simula-tion. The Press-Schechter formalism predicts (for the ellipsoidal collapse description of [71] which best fits the numerical halo abundance data) that about half of the other 40% should be in halos between the simulation resolution limit and $10^{-6}\,M_\odot$, the lower limit set by free-streaming for a 100 GeV WIMP [121]. The remaining 20% of all mass is predicted to be truly diffuse, i.e. never to have been part of any object. A similar calculation at $z = 8$ suggests that just over half of all matter would still be diffuse at that time. Very high resolution simulations of individual halos give insight into whether these large diffuse fractions are reflected in the material accreted during halo build-up. For the six Aquarius halos, 30-40% of their final mass was accreted in objects below the resolution limit of the simulations, $\sim 2 \times 10^5\,M_\odot$ [122]. Extrapolation to unlimited resolution using excursion set theory shows that a typical galaxy halo would accrete about 10% of its mass in diffuse form [121]. Since only halos with $M > 10^8\,M_\odot$ contain even a faint dwarf galaxy, about half of the mass accreted during the growth of the Milky Way's halo was never previously associated with stars.

Mergers are an important component of hierarchical halo growth. However, their importance is often exagger-ated. Major mergers (i.e. those with progenitor mass ratios greater than 1:10) contribute less than 20% to the total mass of the six final Aquarius halos and similar amounts are found for halos over a wide range in mass [120, 122, 123]; the bulk of the material comes, in roughly equal amounts, from minor mergers and from "diffuse" material in which the simulations resolve no structure (some of this material was, in fact, ejected from earlier, resolved halos).

## 5.2 Halo abundance and structure

The abundance and internal structure of halos as a function of their mass and of time has been the subject of intense study since the early days of N-body simulations. DEFW [44] investigated the broad properties of CDM halos (which they called "groups"), identified using the friends-of-friends, or FoF, algorithm in which particles are linked together if they are closer than a certain length. In spite of their small size and poor mass resolution, these simulations outlined the overall shape of the halo mass function showing, in addition, that halos are strongly triaxial and rotate slowly. With higher resolution simulations [49, 52], it became clear that the halo mass function has power-law shape with index close to -2, and a high mass cutoff that grows rapidly with time, especially in high density regions. By implementing the procedure now known as "halo abundance matching," Frenk et al. [52] showed that their CDM simulations roughly reproduced the shape and amplitude of the Tully-Fisher/Faber-Jackson relation, implying a good correspondence between the predicted abundance of dark matter halos and the observed abundance of galaxies. This was an early success of the cold dark matter model that helped to stimulate further detailed investigations.

### 5.2.1 The halo mass function

The halo mass function in the ΛCDM model, and its variation with redshift, have now been estimated with high precision using simulations of different size. For example, Fig. 5 shows that the Millennium series of simulations establishes a numerically converged and statistically well determined halo mass function over almost seven orders of magnitude, from about $10^9 \, M_\odot$ up to nearly $10^{16} \, M_\odot$ [117]. This figure illustrates how such simulations of overlapping resolution can be used to estimate the size of numerical artifacts, for example, the tendency of the FoF algorithm to overestimate the halo mass when halos are resolved with fewer than about 100 particles [124]. For a suitable halo definition, such as that implied by the FoF algorithm, the halo mass function turns out to be almost independent of epoch, of cosmological parameters and of the initial power spectrum, when expressed in appropriate variables. A convenient formula is given by Jenkins et al. [125]:

$$M \frac{\mathrm{d}n}{\mathrm{d}M} = \rho_0 \frac{\mathrm{d}\ln\sigma^{-1}}{\mathrm{d}M} f(\sigma(M)), \tag{2}$$

where $\rho_0$ is the mean mass density of the universe, $\sigma(M)$ the variance of the linear density field within a top-hat filter containing mass $M$ and $f(\sigma)$ is a function that is determined empirically by fitting to the simulations. This universality is of the kind predicted by the Press-Schechter model, although the shape of $f(\sigma)$ found in the simulations differs significantly from the original prediction of this model, agreeing much better with a version motivated by ellipsoidal rather than spherical collapse [71].

In practice, significant deviations from the universal form of Eqn. 2 are apparent in high N-body simulations [124–126]. For example, Tinker et al. [126] find that the amplitude of $f(\sigma)$ decreases monotonically by 20-50%, depending on the mass definition, from $z = 0$ to 2.5 and that the overall shape of the mass function evolves somewhat with redshift. Because of this, if high accuracy is desired, different fitting formulae are required for different redshifts and different halo definitions, even for the standard ΛCDM model. For example, the most recent fit for $f(\sigma)$ at $z = 0$, accurate to 5% for FoF halo masses, is given in [117]:

$$f(\sigma(M)) = 0.201 \times \left[ \frac{2.08}{\sigma(M)} \right]^{1.7} \exp\left[ \frac{-1.172}{\sigma^2(M)} \right]. \tag{3}$$

Useful fitting formulae for $f(\sigma)$ for $z < 2$ and for $z > 10$ have been provided in [126] and [127].

### 5.2.2 Halo density profile

The first CDM simulations capable of resolving the internal structure of dark halos revealed the importance of mergers during their assembly [49, 52]. This suggested an explanation for the Hubble sequence of galaxy morphologies along the lines proposed by Fall [128]: spiral disks would form in relatively isolated halos with quiescent histories where gas could condense directly, while ellipticals would form in more clustered regions where (major and minor) mergers transfer energy and angular momentum from the stars to the dark matter, allowing the stellar merger remnants to be compact and slowly rotating. The circular velocity curves of halos in these early simulations broadly supported this picture: those of relatively quiescent halos were roughly flat, whereas those of the more massive halos that had ex-perienced more mergers were rising in the inner regions[5]. This seemed encouragingly consistent with observations showing that the outer rotation curves of spiral galaxies are typically flat, whereas the stellar velocity dispersion in the central galaxies of groups and clusters is typically lower than the velocity dispersion of group galaxies.

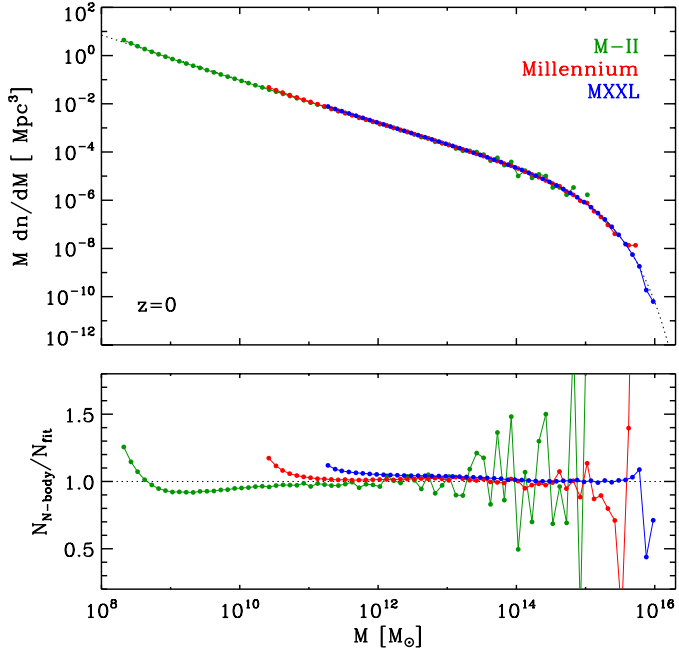

**Figure 5** The differential mass function of dark matter halos. Halos were identified in the Millennium, Millennium-II and Millennium-XXL simulations using a friends-of-friends algorithm with linking length 0.2 times the mean interparticle separation [44]. Each colour corresponds to one of the three simulations. The particle mass varies by a factor of about 1000 between them, from $9.84 \times 10^6 \, \mathrm{M_\odot}$ in the Millennium-II to $8.46 \times 10^9 \, \mathrm{M_\odot}$ in the MXXL and the volume by a factor of 27000, from $(142.86 \, \mathrm{Mpc})^3$ to $(4.11 \mathrm{Gpc})^3$. In all cases the mass functions are truncated at the low-mass end at the mass corresponding to 20 particles. The good agreement between the three simulations indicates that the mass function has converged over the range resolved in each simulation. Together they determine the mass function well over seven orders of magnitude. The bottom panel gives the ratio of the three mass functions to an analytic fitting formula given by Eqns. 2 and 3 and shown as a dotted curve in the top panel. [Adapted from [117]].

The dependence of the shape of circular velocity curves on initial fluctuation power spectrum was systematically investigated in the late 1980s [129,130]. This work showed that they are neither flat, nor well approximated by power laws, but rather steepen gradually from centre to edge. In addition, higher mass halos and halos formed from initial conditions with more large-scale power (i.e. more negative $n$ in $P(k) \propto k^n$) have circular velocity curves which rise for longer in the inner regions, and decline less rapidly at large radius. The non-power law nature of these curves became more evident in higher resolution simulations of the formation of individual halos [131] but was only systematized in the mid-1990's, when it became clear that, to a good approximation, all dark matter halos have circular velocity profiles which are the same shape in a log-log plot, independent of initial fluctuation spectrum, of halo mass, of epoch and of cosmological parameters such as $\Omega_m$ or $\Lambda$ [132,133] (hereafter NFW). This remarkable regularity can be expressed by saying that the spherically averaged density profile of halos can be well fitted by a simple two-parameter formula that has become known as the "NFW profile":

$$\rho(r) = \frac{\rho_s}{(r/r_s)(1+(r/r_s))^2}, \qquad (4)$$

where $r_s$ and $\rho_s$ are a characteristic radius and density, respectively, and are simply related to the radius, $r_{-2}$, at which the logarithmic slope is $-2$, and the density at this radius, $\rho_{-2}$. The mass of a halo is conventionally taken to be the "virial mass", i. e. the mass within the radius, $r_{200}$, which encloses a mean density 200 times the critical value required for closure. At this radius the period of a circular orbit is 63% of $H_o^{-1}$, the Hubble time. Halo concentration can then be defined as $c = r_{200}/r_s = r_{200}/r_{-2}$, and the systematic trends of circular velocity profile with halo mass and initial fluctuation spectrum mentioned above can all be understood as dependences of concentration on these properties[6]. Density profiles with this general form have been shown to arise even in the absence of hierarchical growth, for example from HDM initial conditions [134, 135] or from sharply truncated initial power spectra [136]. The physical origin of this near-universal shape is not well understood, although many attempts at explanation have been put forward [137–142].

---

[5] The circular velocity curve of a halo is defined by $V_c^2(r) = GM(r)/r$, where $M(r)$ is the mass contained within radius $r$. Thus for a spherical halo, $V_c(r)$ gives the orbital speed on a circular orbit at $r$. It has effectively the same information content as the spherically averaged mean density profile, $\rho(r)$, but is less noisy because of its cumulative nature.

NFW showed that the dependences of halo concentration on mass, initial fluctuation spectrum and cosmological parameters all reflect a dependence of concentration on a suitably defined halo formation time. Halos typically assemble earlier, and thus have higher concentration, if they are lower mass, if the cosmic matter density $\Omega_m$ is lower, if the universe is open rather than flat, and if the effective index of the initial fluctuation spectrum $n$ is more positive. Indeed, in a given cosmology the scatter in concentration among halos of given mass correlates strongly with the scatter in formation time. In currently viable $\Lambda$CDM models, the variation of concentration with halo mass is weak, decreasing from about 8 for typical galactic halos, $M \sim 10^{12}\,\mathrm{M}_\odot$, to about 4.5 for typical large cluster halos, $M \sim 10^{15}\,\mathrm{M}_\odot$ [143]. Several fitting formulae have been proposed for this halo-concentration mass relation and its variation with redshift [133, 144–149]. At a given halo mass, however, there is a large scatter in concentration which can be characterized with large statistical samples. Neto et al. [143] find that the dispersion in the logarithm of the concentration is typically $\sim 0.1$ for relaxed halos and can be as high as $\sim 0.15$ for unrelaxed halos. Thus, for example, about 1% of cluster halos ($M_{200}\gtrsim 4 \times 10^{14}\,\mathrm{M}_\odot$) have concentrations exceeding 7.5, and a similar fraction of galactic halos ($M \sim 10^{12}\,\mathrm{M}_\odot$) have concentrations below 4.5.

As more and higher resolution simulations were carried out, small but systematic deviations from the NFW profile became evident [136, 150]. These are particularly convincing when results for many similar halos are averaged in order to suppress the substantial object-to-object variation. To introduce the flexibility needed to account for these differences, Navarro et al. [151] proposed a new 3-parameter fitting formula for the spherically averaged density profile:

$$\ln \rho(r)/\rho_{-2} = (-2/\alpha)\,(r/r_{-2})^{\alpha}. \qquad (5)$$

---

[6] The definition of halo "radius", and hence of halo mass, is arbitrary because halos have no physical edge. The value $r_{200}$ was used in the original NFW work but, unfortunately, not all subsequent authors have adopted the same convention. Readers should be aware of potential confusion introduced by this lack of consistency in definitions of halo mass, radius and concentration.

The adjustable shape parameter, $\alpha$, shows considerable scatter but increases systematically with halo mass at $z = 0$ in the standard $\Lambda$CDM cosmology, reflecting small but real deviations from a truly universal profile [147]. (The average Aquarius galactic halo has $\alpha = 0.159$ while the average Phoenix cluster has $\alpha = 0.175$ [113].) No systematic studies have been carried out so far to test how $\alpha$ varies with initial power spectrum, with cosmological parameters, or even with epoch in the standard $\Lambda$CDM cosmology. Since Equation (5) had previously been used by J. Einasto to fit star counts in the Milky Way [152], this fitting function has come to be known as the "Einasto profile'.'

The highest resolution simulations of individual halos currently represent the mass distribution within $r_{200}$ by more than $10^9$ particles, e.g. the largest simulations in the Virgo Consortium's "Aquarius" and "Phoenix" Projects studying galactic and cluster halos, respectively [112,113], and also the GHALO simulation of a galactic halo [115]. Although only a handful of simulations exist today at this resolution, there are a number with more than $10^8$ particles within $r_{200}$, six as part of the Aquarius Project, nine as part of the Phoenix Project, and the "Via Lactea-II" simulation [114]. In the case of the highest resolution Aquarius simulation and in GHALO, the particle mass is $\approx 1000\,\mathrm{M}_\odot$, more than 6 orders of magnitude lower than in the first halo simulations of Frenk et al. [49] and 5 orders of magnitude improvement over the galaxy halo simulations of NFW!

As far as spherically averaged density profiles are concerned, the new generation of simulations has confirmed the trends uncovered by previous generations. The current situation is summarized in Fig. 6 which shows such profiles for the Aquarius galaxy halos and the Phoenix rich cluster halos. In the left panel, each profile is multiplied by $r^2$ and is scaled to its individual estimated values of $r_{-2}$ and $\rho_{-2}$. The thin lines are results for the individual Phoenix halos, while the thick dashed lines give mean results obtained by stacking all nine Phoenix and all six Aquarius halos. The right panel shows the same data but expressed in terms of the logarithmic slope of the density profiles as a function of scaled radius. Profiles are reliably measured down to 1% of the scale radius, $r_{-2}$, corresponding to about 0.2% of the virial radius, $r_{200}$. In spite of the considerable scatter (which is somewhat larger for clusters than for galaxy halos), the mean Aquarius and Phoenix profiles are close to each other, with Aquarius being somewhat more "peaked", corresponding to larger $\alpha$. The logarithmic slope, $\gamma(r) \equiv d\ln\rho/d\ln r$, near the inner resolution limit is close to $-1$, although in individual Phoenix clusters it can be as small as $-0.7$ or as large as $-1.5$.

There is considerable work analysing the spherically averaged velocity dispersion structure of dark halos in parallel with analysis of their density structure. In a spheri-cal, equilibrium, collisionless system there are two independent profiles, $\sigma_r(r)$ and $\sigma_t(r)$, the *rms* dark matter velocities in the radial and tangential directions, respectively. The function $\beta(r) \equiv 1 - 0.5\sigma_t^2(r)/\sigma_r^2(r)$ is known as the velocity anisotropy. $\beta(r) = 1$ and $\beta(r) \rightarrow -\infty$ correspond to systems where dark matter particles have purely radial and purely circular orbits, respectively. $\beta(r)$ is identically zero in a strongly collisional system like a gas sphere; the velocity dispersion is then said to be everywhere isotropic. Typical simulated dark matter halos turn out to be almost isotropic in their inner regions and to be somewhat radially biased at larger radii. Just as for the density profiles, there is substantial variation from halo to halo, both in $\beta(r)$ and in the shape of the overall velocity dispersion profile, $\sigma^2(r) \equiv \sigma_r^2 + \sigma_t^2$. Results for the Aquarius and Phoenix halos are given in [153] and [113].

Hansen and Moore [154] suggested that halo formation might lead to a general relation between the local value of $\beta(r)$ and the logarithmic slope of the density profile, $\gamma(r)$. The Aquarius halos satisfy this relation quite well in their inner regions, but not at larger radii. A better established, but quite mysterious, regularity was pointed out by Taylor and Navarro [155]. Although both $\rho(r)$ and $\sigma(r)$ vary substantially from halo to halo and neither is close to a power law in any halo, the combination $f(r) \equiv \rho(r)/\sigma^3(r)$, the pseudo-phase-space density, is remarkably close to the same power law, $f \propto r^{-1.875}$ in *all* halos [153]. The reason for this regularity is not currently understood.

### 5.2.3 Halo shape and spin

The triaxial shape of halos was apparent in the earliest cold dark matter simulations [44, 52]. This reflects formation through the anisotropic collapse of ellipsoidal overdensities in the initial mass distribution, and subsequent growth by accretion and mergers. These processes induce clear alignments between halos and the surrounding dark matter density field. The matter within halos ends up stratified on concentric triaxial ellipsoids whose axes are relatively well aligned [157–159]. The joint distribution of axial ratios was first characterized by Jing and Suto [157]. Typical $\Lambda$CDM halos have three unequal axes and there is a weak preference for near-prolate over near-oblate shapes. Major to minor axis ratios exceeding a factor of two are quite common, and there is a weak tendency for more massive halos to be more aspherical.

This work has recently been updated and extended to a larger mass range, $10^{10}$–$2\times10^{14}\,\mathrm{M_\odot}$, using the Millennium-I and II simulations [81, 159], which are also well adapted to studies of alignments within and beyond halos [81, 159, 160]. As already noted in previous studies of smaller sam-ples [161–164], the inner parts of halos tend to be more aspherical than the outer parts; more massive halos tend to be more aspherical than less massive ones; halos at high redshift are more aspherical than their low-redshift counterparts. Although the mean angles between the halo major axes at small and large radii are typically $\sim 20°$, there is very large scatter in the distribution of alignment angles. About 25% of halos have nearly perpendicular major axes at small and large radii. Halos show strong alignments with the surrounding matter distribution out to tens of megaparsecs [159, 165, 166]. These become more prominent with increasing halo mass, reflecting the relative youth of the more massive halos. These systematics of halo shapes and orientations are important for galaxy formation and for both dynamical and gravitational lensing studies.

Halo shapes are supported by anisotropic velocity dispersions, not by rotation. Halos acquire angular momentum through tidal torques [167, 168] but it was discovered very early that they tend to rotate slowly [169]. This may be characterized by the dimensionless spin parameter $\lambda$,

$$\lambda = \frac{J|E|^{1/2}}{GM^{5/2}}, \tag{6}$$

where $M$ is the halo mass, $J$ is the magnitude of its angular momentum vector and $E$ is its total energy. The spin parameter measures the amount of coherent rotation in a system compared to random motions and can be thought of as the ratio of a typical rotation velocity to the *rms* velocity. For a purely rotationally supported self-gravitating object, $\lambda \simeq 0.4$. Halos formed by hierarchical assembly from cosmological initial conditions have a median value of $\lambda \approx 0.04$ with very little dependence on mass or cosmology [130, 161]. The distribution of $\lambda$ for equilibrium FoF halos (for the definition of equilibrium, see [81]) is well fit by a lognormal function,

$$P(\log\lambda) = \frac{1}{\sigma\sqrt{2\pi}} \exp\left[-\frac{\log^2(\lambda/\lambda_0)}{2\sigma^2}\right], \tag{7}$$

with peak location $\lambda_0 = 0.04222 \pm 0.000022$ and width $\sigma = 0.2611 \pm 0.00016$.

In standard $\Lambda$CDM the spin parameter depends only very weakly on halo mass: the most massive halos tend to spin slowest, but the strength of the trend depends on the exact definition of a halo [81]. The total spin of a halo does, however, correlate with its shape: more aspherical halos typically rotate faster. This reflects a correlation between the shape of a protohalo and the strength of its coupling to the tidal field at early times. Most halos have their spin axis roughly aligned with their minor axis and lying perpendicular to their major axis. However, the distribution

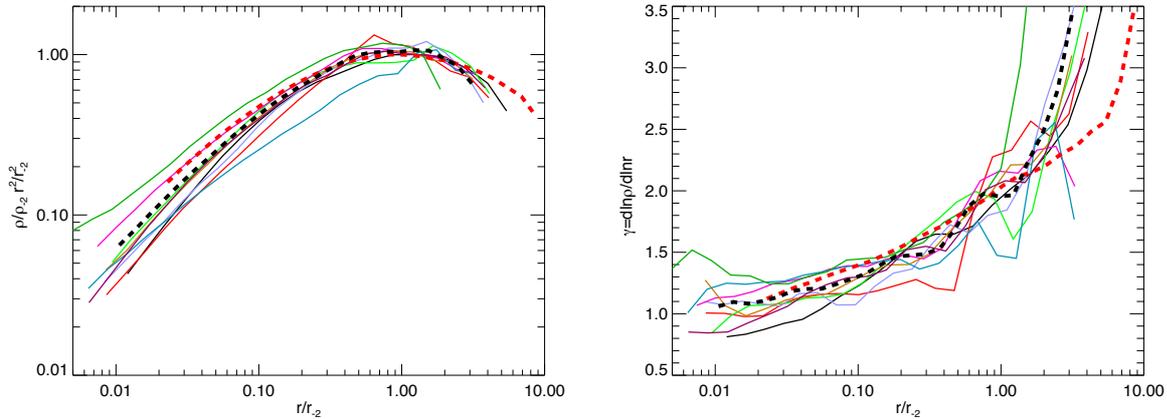

**Figure 6** Spherically-averaged density (left) and logarithmic slope (right) profiles for the nine Phoenix rich cluster halos as a function of radius. Radii are scaled to the characteristic radius, $r_{-2}$, (the radius at which the logarithmic slope has the "isothermal" value of -2 in the best-fit Einasto profile). Profiles are plotted down to the minimum numerically converged radius, $r_{conv}$, defined by [156]. The thick dashed black line shows the mean density profile for a stack of all nine Phoenix halos, made after scaling each to its own virial mass and radius. The thick red dashed line shows the result of the same stacking procedure, but applied to the six Aquarius galaxy halos.

of alignments with respect to all three axes is fairly broad [81, 162, 164, 165, 170].

The spins and shapes of halos are sensitive to their large-scale environment: more rapidly rotating halos of a given mass are more strongly clustered, as are rounder halos, even though asphericity correlates positively with rotation [81]. The strength of these correlations increases with halo mass: it is weak for galactic halos, but can reach a factor of two for galaxy cluster halos. This is a further indication that the internal properties of halos depend not only upon their mass but also upon the environment in which they form.

The internal distribution of angular momentum within halos is typically fairly regular. On average, for halos of a given mass, the median specific angular momentum, $j$, increases with radius as $j(\leq r) \propto r$. Thus, halos do not rotate like solid bodies, but rather have an angular velocity that scales roughly as $r^{-1}$. However, there is a large amount of scatter around these trends [171]. The cumulative distribution of $j$ can be fit by a universal function which follows a power-law, $M(< j) \propto j$, over most of the mass, and flattens at large $j$ [144]. The direction of the angular momentum vector varies considerably with radius: the median angle between the inner ($r \leq 0.25 r_{200}$) and total ($\leq r_{200}$) angular momentum vectors is about 25% [171]. Again there is large scatter: 95% of halos have their total angular momentum directed between 5° and 65° from the inner direction.

The large scatter in the angular momentum structure of halos reflects the stochastic nature of halo assembly. Merg-ers, both major and minor, can have dramatic effects, particularly in the inner parts. For example, analysis of simulated galactic halos [172] shows that large changes in the direction of the angular momentum vector occur frequently: over their lifetimes (i.e. after a halo acquires half of its final mass), over 10% of halos experience a flip of at least 45° in the spin of the entire system and nearly 60% experience a flip this large in the inner regions. Such changes, often associated with misalignments between the shape and angular momentum of halos, can have drastic effects on the properties of the galaxy forming in the halo, sometimes inducing major morphological transformations [173].

## 5.3 Halo substructure

Cold dark matter halos are not smooth: vast numbers of self-bound susbtructures ("subhalos") swarm within them. Subhalo centres are the sites where cluster galaxies or satellites galaxies should reside. Substructures were identified as soon as N-body simulations of halos reached sufficiently high resolution [136, 174]. It was immediately apparent that a significant fraction of the halo mass is tied up in subhalos and that most of them reside in the outer parts of the main halo, those that venture close to the centre being stripped or disrupted by the strong tidal forces. Subhalos have cuspy, NFW-like density profiles but, because of tidal stripping, they tend to be less extended than comparable halos in the field [112,114,174].

Soon after the presence of subhalos was established, their existence was flagged as potentially disastrous for the cold dark matter model because the number seen in simulations of galactic halos vastly exceeds the number of known satellite galaxies orbiting the Milky Way [136, 175]. (By contrast, the number of subhalos in simulations of cluster halos is comparable to the number of cluster galaxies [95, 136, 176].) As we shall see below, a plausible explanation for the discrepancy is that galaxy formation is extremely inefficient in small halos: various forms of feedback (such as reionization or the injection of supernova energy) render all but a handful of the largest subhalos invisible. Although deprived of stars, these dark subhalos are, in principle, detectable from their gravitational lensing effects [177–179].

The large number of subhalos in cold dark matter simulations is evident in the images of galaxy and cluster halos shown in Fig. 4. It is quantified in Fig. 7 which shows the average cumulative number of subhalos in the Aquarius and Phoenix simulations as a function of $\mu = M_{sub}/M_{200}$, the gravitationally bound mass of the subhalo in units of the parent halo mass. Approximately 10,000 subhalos are seen within $r_{200}$ in both cases above the simulation resolution limit. For both galaxy and cluster halos, the mass function is well fit over 5 orders of magnitude in fractional mass by a power-law, $N(> \mu) \propto \mu^s$, with $s = 0.94 \pm 0.02$ for Aquarius and $s = 0.97 \pm 0.02$ for Phoenix. These exponents are very close to the critical value of unity, for which each logarithmic mass bin contributes equally to the total mass in substructure. This is logarithmically divergent as $\mu$ approaches zero, and implies that a significant fraction of the mass could, in principle, be locked in halos too small to be resolved by the simulations. However, even at the resolution of the largest Phoenix simulation, spanning seven decades in subhalo mass, only 8% of the mass within $r_{200}$ is in substructure. Extrapolating down to the Earth mass assuming $N \propto \mu^{-1}$, the total mass in substructure is predicted to be about 27%.

As may be seen in Fig. 7, at a given value of $\mu$, there are somewhat more subhalos in cluster than in galactic halos. Indeed the mean resolved substructure mass fractions within $r_{200}$ are 7% for the Aquarius halos but 11% for the Phoenix halos (down to $\mu \sim 10^{-7}$ in both cases), with a range of about a factor of 1.5 for the 9 Phoenix simulations [113]. This shift reflects a genuine difference between galaxy and cluster halos arising from their relative dynamical age: substructure is more effectively destroyed by tides in the older, galactic halos.

The abundance of subhalos also varies systematically with other properties of the parent halo, although the scat-ter is large. For example, at fixed halo mass, the amount of substructure decreases with halo concentration and with halo formation redshift [176]. On the other hand, there are properties of the subhalo population that depend weakly, if at all, on the properties of the parent halo. For example, their radial distribution varies little with the mass (or concentration) of the parent halo; it is much less centrally concentrated than the overall dark matter profile, and, surprisingly, is independent of subhalo mass [112]. (This does depend somewhat on the way in which subhalos are defined [180].)

Although the subhalo mass functions shown in Fig. 7 depend on parent halo mass, it is remarkable that the abundance of low-mass subhalos *per unit parent halo mass* is nearly independent of parent halo mass. In fact, it is similar to the abundance per unit mass of low-mass halos in the Universe as a whole, once differing boundary definitions for subhalos and halos are accounted for [176]. This is due in part to the relatively recent accretion epoch of surviving subhalos: about 90% per cent of present-day subhalos were accreted after $z = 1$ and about 70 per cent after $z = 0.5$, almost independently of subhalo or parent halo mass [176]. Survival in the strong tidal fields of the parent halo is tough: only about 8% of the total halo mass accreted at $z \sim 1$ survives as bound subhalos at $z = 0$ and most of these subhalos are at relatively large radii. In the inner regions where the Sun resides, the fraction of halo mass in subhalos is much smaller than the average value within $r_{200}$. As we see below, this has major consequences for detection experiments.

When comparing with observed satellite galaxies the maximum circular velocity of a subhalo is a much more relevant property than its mass, since it is much less affected by tidal stripping and can be inferred (with some extrapolation) from the dynamics of the stars and gas. It turns out that when $V_{max}$ is expressed in terms of the circular velocity of the parent halo, $V_{200}$, the abundance of subhalos is nearly independent of $V_{200}$. That is, if $\nu$ is defined as $V_{max}/V_{200}$, then $N(> \nu)$ is another simple, nearly universal function describing (sub)halo properties, with $N(> \nu) = 10.2(\nu/0.15)^{-3.11}$ (over the range $0.05 < \nu < 0.5$) [181].

To summarize this whole section, if baryonic effects on halo structure are neglected, cold dark matter halos have the following properties:

– They form hierarchically through mergers of objects with a very wide range of mass, including a significant diffuse component. Most mergers are minor and deposit their mass at large radius; major mergers bring material to halo centre but are subdominant. Formation is thus predominantly "inside out".

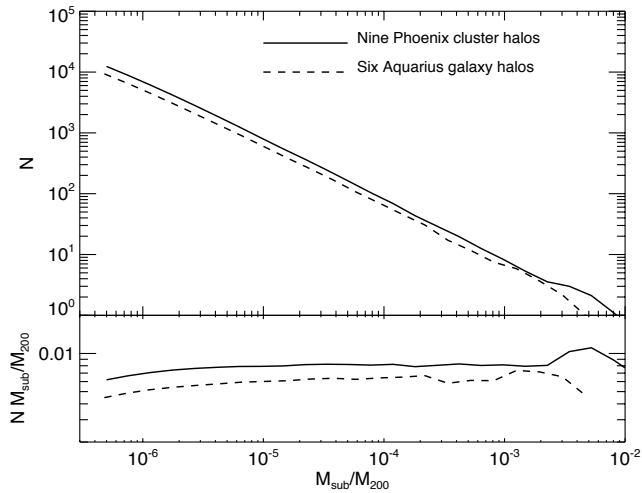

**Figure 7** The cumulative abundance of subhalos within the radius $r_{200}$ in cluster and galaxy halos. The solid lines show the mean subhalo mass functions for the nine Phoenix [113] cluster halos and the dashed lines for the six Aquarius galaxy halos [112]. The bottom panel shows the same data as the top panel, but with the number of subhalos multiplied by mass.

– The halo mass function, when expressed in terms of suitable variables, has an approximately universal form, independent of epoch, cosmological parameters and initial spectrum of density fluctuations.

– The spherically averaged density profile of halos also has an approximately universal form. This is described to $\sim 10 - 20\%$ accuracy by the NFW formula, and to even higher accuracy (in the mean and at the expense of an additional parameter) by the Einasto profile.

– Halo density profiles are "cuspy", i.e. the density diverges towards the centre with a logarithmic slope similar to, or even flatter than $-1$ at the smallest numerically resolved radii.

– Halo shapes are generically triaxial, reflecting the highly anisotropic nature of halo collapse, which also induces strong alignments with the surrounding matter distribution.

– Halos rotate slowly, with a typical spin parameter $\lambda \simeq 0.04$, which depends weakly on mass and also on local environment density. The angular momentum structure is fairly regular but halos often experience spin flips, particularly in their inner parts.

– Bound subhalos with $\mu = M_{sub}/M_{200} > 10^{-7}$ contain about 10% of the halo mass within $r_{200}$, but much smaller fractions in the inner regions where the central galaxies lie. Lower mass halos tend to have fewer subhalos and lower subhalo mass fractions at given $\mu$. Subhalos have cuspy profiles with circular velocity curves peaking at smaller radii than for isolated halos of the same $V_{max}$.

– The cumulative subhalo mass function is a power-law $N(>\mu) \propto \mu^{-s}$ for $\mu \ll 1$, with $s$ close to the critical value of unity. The largest simulations to date indicate that $s$ is slightly less than one implying that a substantial fraction of subhalo mass is contained in a relatively small number of massive subhalos.

– The subhalo population is accreted relatively recently and is much less centrally concentrated than the dark matter: subhalos tend to reside in the outer parts of their parent halos. This radial distribution appears independent of subhalo mass.

– The population of subhalos can be conveniently characterized by the distribution of $\nu = V_{max}/V_{200}$, which is very nearly independent of the parent $V_{200}$.

– There are large halo-to-halo variations in most properties of halos and their subhalo populations.

These regularities are of fundamental importance for studies of galaxy formation and also for attempts to detect particle dark matter directly or indirectly. It is important to bear in mind, however, that many properties of halos and subhalos are modified by the effects of baryons cooling within them. Some of these effects are discussed in § 7 below.

## 6 Dark matter detection: fine-scale structure in the dark matter distribution

The idea that dark matter is made of free elementary particles of some new type will likely be fully accepted by the scientific community only once non-gravitational effects of these particles have been unambiguously measured. Two routes to making such measurements are being actively pursued. Direct detection experiments attempt to measure the effects of dark matter particles in a laboratory. The mean dark matter density in the Solar neighbourhood, which is estimated from the kinematics of nearby stars, is

~ 0.3 Gev/cm$^3$ [182, 183], so for DM particles of mass $m_{GeV}$, the flux through a detector is ~ $10^7/m_{GeV}$ cm$^{-2}$s$^{-1}$. Even for the very small interaction cross-sections expected for supersymmetric WIMPS, collision rates within a detector are measurable provided backgrounds can be controlled. For axions, detection is feasible by looking for resonant interactions with the photons in a highly tuned microwave cavity. A critical issue is whether the dark matter density seen by such experiments should be close to the estimated mean or could differ substantially from it, as expected, for example, for a small-scale fractal distribution, or if most DM were bound into small-mass clumps. Most experiments are sensitive to DM particle energies or momenta so the expected local velocity distribution is also of interest: for example, could a significant fraction of the signal come from a stream of dark matter with locally well defined velocity, produced, perhaps, by the tidal disruption of an infalling subhalo?

The second route to non-gravitational detection is by searching for annihilation or decay products from distant dark matter concentrations. In typical supersymmetric models WIMPS can decay and can annihilate with each other. Both processes can have a significant branching ratio into photons which would then be detectable with $\gamma$-ray telescopes. Since particle decay is independent of environment, it would produce $\gamma$-ray images of dark matter halos which trace their mass distribution, just as optical images of galaxies trace their star distribution. The situation is more complex for annihilation radiation, however, because annihilation probability is proportional to the local density of dark matter and so is enhanced in dense regions. For example, for an NFW model of concentration 10, which might be a good representation of the smooth part of the Milky Way's halo, half of the decay radiation within the virial radius $r_{200}$ ~ 250 kpc would be generated within 0.36$r_{200}$, whereas half of the annihilation radiation would be generated within 0.026$r_{200}$, i.e. within the Sun's orbit around the Galactic Centre [118].

One consequence of this density dependence is that annihilation from CDM halos is expected to be dominated by the contribution from low-mass subhalos rather than from smoothly distributed dark matter. Although such subhalos are projected to contain at most a few percent of the halo mass, their high characteristic density more than compensates. The exact factor by which emission from a halo is "boosted" by this effect is a strong function of radius (reflecting the radial dependence of the subhalo mass fraction) and is quite uncertain because it depends on extrapolating the abundance and concentration of subhalos to masses much smaller than can be reached by direct simulation [118, 184].

Fig. 8 shows results based on the Aquarius A-1 halo, one of the largest simulations of a "Milky Way" halo to date. The annihilation from the smooth DM component is strongly concentrated towards the Galactic Centre on the (simulated) sky, whereas emission from resolved subhalos (i.e. with masses > $10^{-7}$ that of the main halo) comes from many objects at apparently random positions on the sky. The dominant contribution, however, (by a factor of two) is projected to come from subhalos with masses in the 11 orders of magnitude between the resolution limit of the simulation and the mass of the Earth, the limit assumed here to be imposed by early thermal motions of the DM particles. The emission from both resolved and unresolved subhalos is almost uniform on the sky because most of these subhalos are much farther from the Galactic centre than the Sun. In consequence, low-mass subhalos dominate the Milky Way's emission as seen by a distant observer by a much larger factor, 230 in this model. This effect is even stronger for more massive objects. Applying similar extrapolations to simulations of rich galaxy clusters, low-mass subhalos are found to dominate their annihilation radiation by more than a factor of 1000 and to result in much broader images than would correspond to NFW fits to the overall mass distribution [113, 185].

In addition to gravitationally bound subhalos, dark matter halos contain other kinds of substructure which may be relevant for dark matter detection. When smaller dark matter objects fall into a halo, much of their dark matter is typically removed by tidal effects. This material continues to follow orbits very close to that of the infalling system, producing tidal streams which are directly analogous to the meteor streams formed along the orbits of disrupted comets. If the Earth were to pass through a massive stream of this kind, a significant fraction of detector events could correspond to dark matter particles with very similar (vector) velocities, just as the meteors in a shower all appear to radiate from the same point on the sky. Such streams can be identified in high-resolution simulations by algorithms which trace the provenance of particles or which search for overdensities in the full 6-D position-velocity phase-space. At positions corresponding to that of the Sun in the Milky Way, 6-D structure-finders applied to the largest currently available simulations find about ten times as much mass in the form of streams as in bound subhalos, but this is still only about one percent of the local DM mass density [186]. Such streams are unlikely to be important for detection experiments unless these are particularly sensitive to the highest velocity particles.

A different kind of stream is a direct consequence of the CDM initial conditions. At early times, after the DM has decoupled from other particle species and has become non-relativistic, but before nonlinear structure has started

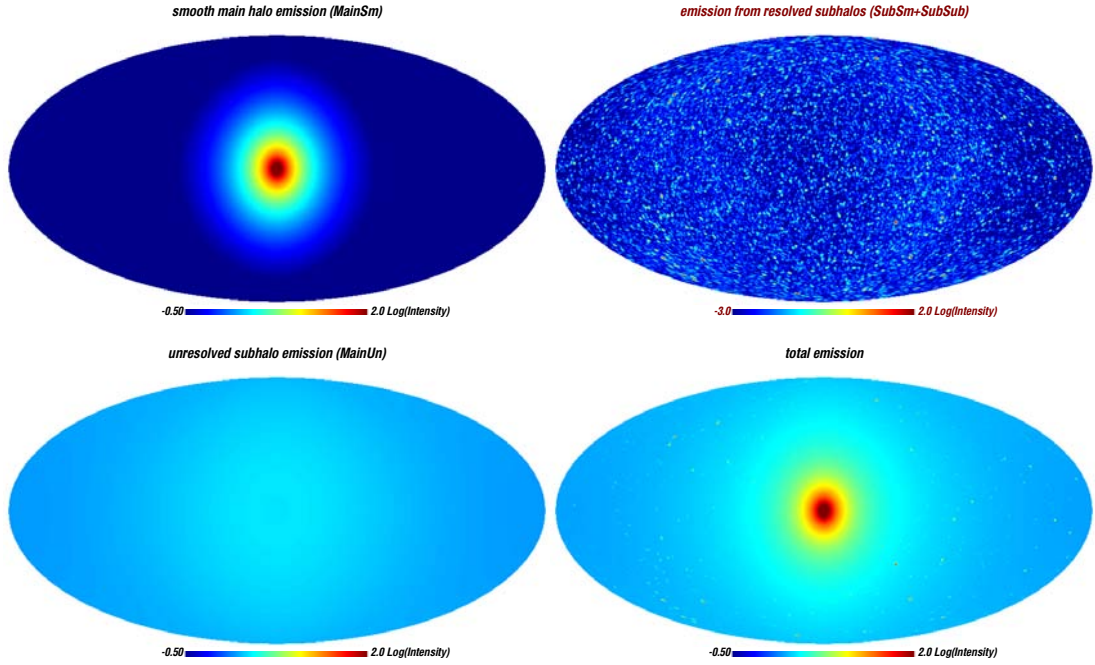

**smooth main halo emission (MainSm)**

**emission from resolved subhalos (SubSm+SubSub)**

**unresolved subhalo emission (MainUn)**

**total emission**

**Figure 8** All-sky maps of the predicted gamma-ray annihilation radiation as seen from the Sun from different components in the highest resolution Aquarius halo (Aq-A-1). The top left panel shows the main halo's diffuse emission while the top right panel shows the emission from all resolved subhalos. The luminosities assigned to each subhalo include the predicted contribution from unresolved (sub-)substructure. For simplicity and for better graphical reproduction they have been represented as point sources smoothed with a Gaussian beam of 40 arcmin. The bottom left panel gives the expected surface brightness from unresolved subhalos down to the free streaming limit. This is a very smooth component that dominates the total flux. Finally, the bottom right panel shows the total surface brightness from all components together. All maps show the surface brightness in units of the main halo's diffuse emission, and use the same color scale, except for the map of the resolved substructures, where the scale extends to considerably fainter surface brightness. [Adapted from [118].]

to form (e.g. at $z \sim 1000$), the phase-space distribution of the DM can be written as:

$$f(\mathbf{x}, \mathbf{v}, t) = \rho(t)[1 + \delta(\mathbf{x}, t)] N([\mathbf{v} - \mathbf{V}(\mathbf{x}, t)]/\sigma(t)), \tag{8}$$

where $\rho(t) \propto (1 + z)^3$ is the mean cosmic density of DM, $\delta(\mathbf{x}, t)$ and $\mathbf{V}(\mathbf{x}, t)$ are the local overdensity and the mean streaming velocity of DM at position $\mathbf{x}$ and time $t$, $N$ is a 3-D unit normal distribution and $\sigma(t) \propto 1 + z$ is the thermal velocity dispersion of the DM particles (around 1 mm/s today for a standard supersymmetric WIMP). The streaming velocity $\mathbf{V}$ is generated by the gravitational effects of the overdensity $\delta$, and we expect $\langle \delta^2 \rangle \ll 1$ and $\sigma^2 \ll \langle \mathbf{V}|^2 \rangle$. The dark matter is thus confined to the neighborhood of a 3-D sheet in the full 6-D phase space and the projection of this sheet onto $\mathbf{x}$-space is nearly uniform.

At later times, evolution of the dark matter is gravitationally driven and effectively collisionless, so it satisfies the collisionless Boltzmann equation $Df/Dt = 0$, where the convective derivative follows the 6-D flow. Thus, at all times, even the present, $f$ is non-zero only in the neighborhood of a 3-D sheet and must take the same values as at early times. The projection of this sheet onto $\mathbf{x}$-space can be stretched and folded but it cannot be torn, so the sheet always passes through every position $\mathbf{x}$, implying there is dark matter everywhere, even outside halos. Inside halos, the sheet is folded by orbital motions and so is typically projected many times onto each $\mathbf{x}$. Within the sheet, particles have very low velocity dispersion ($\sim \sigma$) so the velocity distribution at some position within a halo (say, the position of a DM detector) is a superposition of discrete streams corresponding to the different parts of the sheet projected onto that position. The degree of structure in the velocity distribution at the detector thus depends on the number of such streams, i.e. the degree of folding of the sheet. At a fold, the sheet is parallel to the **v**-direction and so projects to a caustic in **x**-space. In the limit $\sigma \to 0$, the annihilation probability approaches unity as particles pass through caustics [187]. Since every particle passes several caustics in each orbit around its halo, it appears that caustics might be a major contributor to the total annihilation luminosity.

The detailed evolution of the phase-sheet in the neighborhood of every particle in an N-body simulation can be followed explicitly by integrating the geodesic deviation equation (GDE) in tandem with the equations of motion [188]. All the caustic crossings of every simulation particle can then be identified, and at any given time, the phase-sheet properties at the positions of these particles (projected 3D density, internal velocity dispersion, mean motion) provide a dense Monte Carlo sampling (by mass) of the simulated region. For small but finite initial $\sigma$, caustic density singularities are broadened; their annihilation luminosity then becomes finite and can be calculated during a simulation by integrating phase-sheet properties along individual particle trajectories [189].

GDE simulations of $\Lambda$CDM galaxy halos show that, for the $\sigma$ values appropriate for supersymmetric WIMPS, annihilation from caustics provides less than 0.1 percent of the signal near the Sun, and will be detectable above other annihilation sources only near the outermost caustic of external systems like the Andromeda galaxy [190]. Near the Sun's position, folding of the phase-sheet produces a very large number of streams – the $\sim 10^6$ densest streams typically account for about half the local density of dark matter. As a result, the velocity distribution relevant for DM detectors is predicted to be very smooth, to a good approximation a trivariate Gaussian [158]. The single densest stream at the Sun is expected to account for about $10^{-3}$ of the local DM density. The velocity dispersion within this stream is $\sim 1$ cm/s, so it would show up in an axion detector as a "spectral line" of width $\Delta E/E \sim 10^{-7}$ containing $\sim 10^{-3}$ of the total flux; this might be detectable [190]. Although the velocity distribution of DM particles is predicted to be quite smooth, their energy distribution is predicted to show broad features which are fossil relics of the detailed assembly history of the Milky Way's halo [158,191].

# 7 Baryonic effects on the dark matter distribution

Our discussion of the development of dark matter structures has so far largely ignored the baryonic material.

Baryons account for about one sixth of all cosmic matter, so this assumption should clearly be looked at carefully. After reionization of the universe by the first galaxies and their active nuclei, photoionization effects set the temperature of the bulk of the baryons to a few thousand degrees. This corresponds to a baryonic Jeans mass similar to the stellar mass of a dwarf galaxy, so baryons are expected to react gravitationally in the same way as the dark matter and hence to follow its distribution on all larger scales, at least until they collapse to make nonlinear objects and additional baryonic processes intervene. This can be tested directly by looking at Ly-$\alpha$ absorption due to diffuse intergalactic hydrogen along the line-of-sight to distant quasars, the so-called Ly-$\alpha$ forest [192]. Recent observations do indeed find structure in the Ly-$\alpha$ forest that is very similar to the predictions of detailed $\Lambda$CDM+baryon simulations all the way down to the limit imposed by the baryonic Jeans mass [193]. This not only provides a quantitative check of the $\Lambda$CDM paradigm down to scales well below those of bright galaxies, it also provides a strong lower limit on possible masses for WDM particles [194]. These are required to be sufficiently massive that predictions for halo properties differ from CDM only for low-mass dwarf galaxies.

As more massive halos collapse, the associated baryons shock, cool and condense at halo centre to form galaxies. Radiative and hydrodynamical feedback from stars and active nuclei in these galaxies can heat the remaining gas, perhaps expelling it from the halos altogether. The most massive halos in today's universe, rich clusters of galaxies, seem to have retained all their associated baryons, mostly in the form of hot X-ray emitting gas. Less massive clusters and groups of galaxies have lower baryon fractions, however, suggesting that increasing amounts of baryons are expelled from lower escape velocity systems [195]. The baryons in halos like that of the Milky Way appear to be predominantly in stars, which account for only about 20% of the baryons expected; most appear to have been expelled to large radius. Thus, except for the most massive systems, halos appear to have lost significant mass relative to the case where baryons and dark matter are similarly distributed. This not only reduces halo masses but also causes the remaining material to expand; typically 10% mass loss results in a 10% increase in the linear size of a halo. Together, these effects imply that $\Lambda$CDM simulations which ignore baryon effects will significantly *over*predict the effects of gravitational lensing on scales where these are dominated by individual halos [196].

Although baryons are substantially underrepresented relative to the cosmic mean when halos are considered as a whole, near halo centre the opposite is usually the case. The visible regions of many bright galaxies appear to be dominated by baryons. In a simple model, the condensation of gas to make the galaxy leads to an adiabatic compression of the inner parts of the dark matter halo, so that its mean density within the final galaxy is substantially larger than in the absence of baryonic effects [197, 198]. Simulations of ΛCDM galaxy formation show considerably more complex assembly paths than this model assumes, however, and suggest that dark matter compression effects are significantly weaker than it predicts, perhaps even absent [199, 200]. This issue will need to be revisited as more realistic galaxy formation simulations become available.

The sudden expulsion of the material from the inner galaxy can affect dark matter profiles in other ways. Navarro, Eke and Frenk [201] showed that if a DM halo grows a relatively massive and compact central disk which is then suddenly removed, the initial compression of its halo is more than reversed, and the final halo can show a constant density core even though the initial halo had a central cusp. They suggested that large amounts of material might be blown out of dwarf irregular galaxies early in their history, thus explaining why many appear to have a core in their dark matter distribution. Later work suggested that for plausible parameters of the initial system, a single violent event of this kind might not produce a big enough effect [202,203]. However, Pontzen and Governato [204] showed that a less violent but repeating variant of the mechanism does indeed turn cusps into cores in recent ΛCDM simulations of dwarf galaxy formation [205]; (see also [206,207]). The material does not need to be removed from the forming galaxy, but only from the (assumed dense) star-forming region, and it can then recondense later to repeat the process. By recycling the same gas many times, the inner DM distribution is sufficiently heated to turn the cusp into a core. This suggests that the rotation curves of dwarf irregular galaxies may no longer be a major challenge to the CDM paradigm.

# 8 Looking to the future

The evidence that the universe conforms to the expectations of the CDM model is compelling but not decisive. Current observational tests span a very wide range of scales. Fluctuations in the microwave background probe from the whole observable universe down to the scale of rich galaxy clusters. Galaxy and quasar clustering probe scales between one and a few hundred megaparsecs. Gravitational lensing measurements constrain the detailed structure of galaxy and cluster halos. Observations of the Ly-α forest at $z \sim 2$ probe down to scales far below those responsible for forming bright galaxies like the Milky Way, putting severe constraints on warm dark matter and on any subdominant contribution from massive neutrinos [194, 208, 209]. Nevertheless, new data are needed to test the model on even smaller scales. Currently, warm dark matter remains possible if the particles are sufficiently massive to evade the Ly-α forest constraints. More exotic possibilities such as self-interacting dark matter may also be possible, provided their properties are carefully tuned. Experimental searches for such alternative types of dark matter would need to be quite different than those for WIMPS or axions, so it is imperative that astrophysicists continue to test the standard model and to evaluate possible evidence for alternatives. Dwarf galaxies are a prime source of such evidence, both in the field and as satellites of larger galaxies.

On subgalactic scales, currently viable hypotheses for the dark matter can produce strikingly different structure. Fig. 9 compares a cold to a warm galactic dark matter halo. The CDM halo is part of the Aquarius simulation series (Aq-A); the WDM halo simulation assumed the same phases for the initial fluctuations, but modified the power spectrum as in Fig. 1 with a cut-off appropriate to a resonantly produced 2keV sterile neutrino [194]. This corresponds approximately to the minimum particle mass (and hence the maximum free-streaming length) allowed by current Ly-α forest data. The main halo is very similar in the two cases with almost indistinguishable mass, flattening, spin and density profile [210]. However, the WDM halo has far fewer substructures (only about 3% of the mass within $r_{200}$ is in substructures, compared to about 8% in the CDM case). In addition, there are more subtle differences, not apparent in the figure. In particular, subhalos within the WDM halo are systematically less concentrated than their CDM counterparts.

There are three aspects of small-scale structure where potential conflicts with the cold dark matter model have been identified: (i) the luminosity function of galactic satellites, (ii) the abundance of galactic substructures as a function of mass or circular velocity and (iii) the structure of the halos that host faint satellites or field dwarfs. Problem (ii) was the first to be clearly identified in the late 1990s when N-body simulations revealed a large number of subhalos within galactic halos, vastly exceeding the number of satellites then known to be orbiting in the halo of the Milky Way [136, 175]. In its original form, the problem was expressed as an excess of subhalos as a function of their circular velocity or mass but it was soon re-expressed as an excess of subhalos as a function of their galaxy luminosity. Although closely related, these two statements are conceptually distinct and the solutions that have been proposed differ for the two cases.

The small number of visible satellites in the Milky Way compared to the large number of subhalos in N-body simu-lations of galactic halos is often referred to as "the satellite problem". Soon after it was first formulated, a new population of ultra-faint dwarf satellites was discovered in the Milky Way [211–215], but they are too small and too few to alter the argument. Possible resolutions to the discrepancy were soon proposed, based on old ideas about galaxy formation. It had been known since the 1970s that at least two processes make it difficult for a visible galaxy to form in a low-mass halo [7, 216]. One is the early reionization of intergalactic hydrogen which raised the entropy of the gas, preventing it from condensing in halos with characteristic velocity below $\sim 10$ km s$^{-1}$. The other is the injection of thermal and kinetic energy into the surrounding gas by supernovae and young stars. A handful of such objects suffice to expel remaining galactic gas in a wind, preventing the formation of subsequent generations of stars. Detailed semi-analytical modelling of these processes developed this as a plausible explanation for the dearth of visible satellites in the Milky Way [90, 217–219]; lower mass subhalos simply do not contain any visible stars. Subsequent exploration of these ideas using hydrodynamic simulations confirmed this conclusion [149, 220, 221], but it must be borne in mind that although it is clear that the intergalactic medium has been photoionized since at least $z = 6$, observational evidence for sufficiently strong winds from dwarf galaxies is more ambiguous.

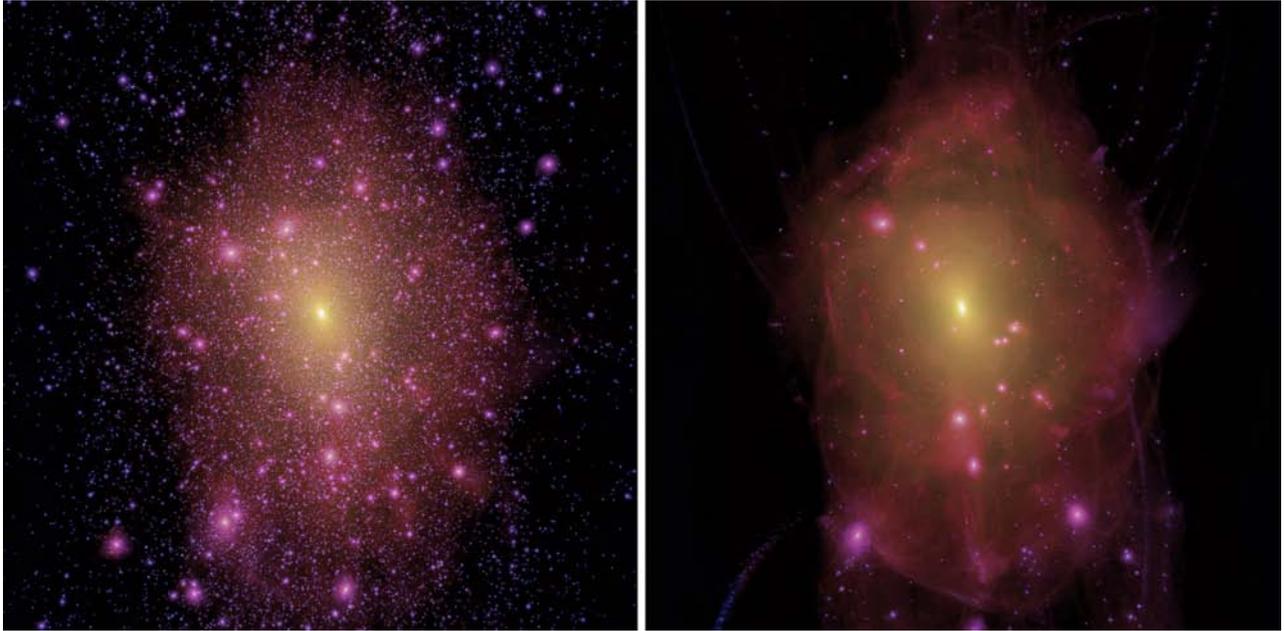

**Figure 9** Images of a CDM (left) and a WDM (right) galactic halo at $z = 0$. The CDM halo is one of the Aquarius simulations (Aq-A); the WDM halo is its warm dark matter counterpart: the fluctuation spectrum in the initial conditions had the same phases, but the power spectrum is truncated as appropriate for a resonantly produced 2keV sterile neutrino. Image intensity indicates the line-of-sight projected density squared, and hue the projected density-weighted velocity dispersion, ranging from blue (low velocity dispersion) to yellow (high velocity dispersion). Each box is 1.5 Mpc on a side. Note the sharp caustics visible at large radii in the WDM image, several of which are also present, although less well defined, in the CDM case.

Problem (ii) cannot be explained away by invoking galaxy formation processes that render subhalos invisible. This is a dynamical problem involving the galaxies that we do see, and whose solution, if it exists, requires either a reinterpretation of the data [218, 222] or a modification of the structure of subhalos. In its clearest form [210, 223], the problem contrasts the relatively low values of $V_{max}$ or concentration inferred from dynamical studies of Milky Way satellites [224–226] with the larger values of $V_{max}$ and concentration measured for subhalos in the Aquarius and via Lactea simulations and for isolated halos in cosmological simulations. A closely related discrepancy is that, at given abundance, field dwarf galaxies have substantially lower $V_{max}$ values than predicted by cosmological simulations [227–231].

Three possibilities have been put forward to explain these discrepancies. One is that the dark matter is not made up of cold particles, but rather of warm particles such as sterile neutrinos [210]. In this case, if the warm particle mass is close to its currently allowed lower limit, the concentration of subhalos is reduced relative to the cold dark matter case by virtue of their later formation epoch. This has the virtue that it also solves the satellite problem in what might reasonably be deemed the most straightforward way: small subhalos are simply not there in the first place.

The other two suggestions are in the context of cold dark matter models. The number of massive subhalos with, say, $V_{max} > 30 \mathrm{km\ s^{-1}}$ increases rapidly with the mass of the parent halo. Thus the discrepancy with the satellite data is reduced if the halo of the Milky Way is significantly less massive than the Aquarius and via Lactea halos where the problem was first identified [223]. This would not, of course, solve the corresponding discrepancy for field dwarfs. The final suggestion is that the baryonic effects discussed at the end of §7 alter the concentration not only of field irregular galaxies, but also of the Milky Way satellites. The uncertainty here is whether this process could work without leaving a more massive long-lived stellar population than is observed in the fainter dwarfs.

Problem (iii) above concerns reported measurements of core radii in dwarf and larger galaxies. This is a long-standing controversy which has seen many twists and turns over the years and is accompanied by a vast literature. Cores have been claimed, for example, for the satellites of the Milky Way [232], for field dwarfs such as DDO154 [233], for low surface brightness galaxies [234] and even for bright galaxies like the Milky Way [235]. Salucci et al. [236] argue in favour of the core radius increasing systematically with the mass of the galaxy. Although it seems clear that consensus has not been reached on the reality or otherwise of cores in dark matter halos, it is worth asking whether cores can be produced in any of the dark matter models currently under discussion.

Contrary to claims in the literature, warm dark matter is not a viable explanation for the sort of cores claimed, for example, by Salucci et al. [236]. Warm dark matter particles obey a phase-space constraint [237] which results in core radii that are smaller for galaxies with larger velocity dispersions [238]. This is exactly the opposite of the trend claimed from the observations. Even for the satellites of the Milky Way, the core radii implied by acceptable candidates for warm dark matter are much smaller than those claimed from the data [239]. If cores are actually present at the centres of halos, then the best bet at present is that they result from the sort of baryonic processes discussed at the end of §7.

An alternative explanation for cores in satellite galaxies is that the dark matter is made up of self-interacting particles [240, 241]. If the self-interaction cross-section increases at low velocities, then recent simulations [242] show that core radii can be induced in small halos like those of the Milky Way satellites without violating the observational constraints from rich clusters that resulted in the exclusion of simpler models [243]. However, such variable cross-section models require considerable fine-tuning to do the job, and, as with WDM, this kind of dark matter cannot explain the claimed scaling of core radius with galaxy size.

The three problems mentioned at the beginning of this section remain a subject of lively debate and suggest clear avenues for astrophysical progress, both theoretical and observational. Higher resolution N-body simulations will further explore predictions for warm and self-interacting dark matter. More realistic N-body and hydrodynamic simulations should clarify whether baryonic effects are indeed viable explanations for the apparent discrepancies with predicted halo structure. Observationally, the statistics of flux ratio and image structure anomalies in strongly gravitationally lensed quasars and galaxies are sensitive to small-scale substructure and so should allow cold and warm dark matter to be distinguished [178, 244, 245]. (See also the recent discussion by Xu and collaborators based specifically on the Aquarius simulations [246, 247].) Indeed, detailed analysis of particularly promising cases has already led to the detection of low-mass dark halos without visible stars [248, 249]. The high redshift universe may offer another opportunity to distinguish between the two models, given the significantly later epoch at which structure is predicted to form in the WDM case and the difficulty in achieving reionization already for standard ΛCDM [250].

# 9 Epilogue

The past thirty years are often referred to as "the golden age" of cosmology. During this period, physical cosmology has emerged as a scientific discipline in its own right, with its own methodology, its own basic premises and with well-defined goals. This period has seen plausible, physically motivated proposals for the geometry of our Universe, for the initial conditions for the formation of all cosmic structures and for the nature of the dark matter. These proposals are all experimentally testable and two of them - the universal geometry and the cosmogonic initial conditions - have been validated experimentally through measurement of fluctuations in the temperature of the microwave background radiation and through extensive surveys of cosmic large-scale structure. The third proposal, that the dark matter consists of non-baryonic, weakly interacting elementary particles, still awaits independent comfirmation.

The past decade also delivered a completely unanticipated discovery: the accelerated expansion of the Universe. While identifying the physical cause for this phenomenon remains a distant aspiration (the label "dark energy" encapsulates our helplessness), the prospects for discovering the dark matter particles seem more promising. Currently favoured candidates plausibly have properties which would allow them to be detected directly in the laboratory, or indirectly through decay or annihilation radiation. In addition, the LHC may unmask indirect particle physics evidence for their existence. Thus there is good reason to hope that this important boundary to our knowledge of the Universe will soon be breached. The techniques and resources are in place for a discovery.

Most searches are focused on cold dark matter. This is entirely justified: the cold dark matter model has been the catalyst for the momentous developments of the past thirty years. The statistical properties of the large-scale distribution of galaxies, the form and amplitude of the microwave background temperature fluctuations and many aspects of galaxy formation were predicted (in advance!) by theoretical models that had cold dark matter at their heart. Time and again when astronomical data seemed to conflict with the model (for example, the clustering of galaxy clusters, large-scale streaming data, galaxies at very high redshift) the conflict has been resolved either by revisions of the data or by refinments of the theory. Nevertheless, even the most ardent supporters of the cold dark matter idea must remain sceptical until the dark matter particles are finally and conclusively discovered.

## Acknowledgements

We would like to thank our many collaborators, particularly students, postdocs and other members of the Virgo Consortium who, over the years, have enriched our own research and served as a source of inspiration. Special thanks to Volker Springel, the author of many of the simulations discussed in the article, and without whose persuasion and encouragement this review would have never been written. Special thanks also to Gao Liang for help producing Figures 4 and 6 and to Raul Angulo for help with Figure 5. We acknowledge the hospitality of the Aspen Center for Physics, supported by NSF Grant 1066293, where the bulk of this article was written. CSF acknowledges a Royal Society Wolfson Research Merit Award and ERC Advanced Investigator grant 267291 COSMIWAY. SDMW acknowledges support from ERC Advanced Investigator grant 246797 GALFORMOD. This work was also supported in part by an STFC rolling grant to the Institute for Computational Cosmology of Durham University and made use of the Cosmology Machine at Durham, part of the DiRAC facility.

**Key words.** dark matter – cosmic structure – simulations